\documentclass{llncs}

\usepackage{amsmath}
\usepackage{amssymb}
\usepackage{stmaryrd} 
\usepackage{cite}
\usepackage{graphicx}
\usepackage{enumerate}
\usepackage{url}
\usepackage{mathrsfs}
\usepackage{eucal}
\usepackage{ifsym}
\usepackage{tikz}

\usepackage[frozencache]{minted}
\setminted{fontsize=\small}
\title{Formal Adventures in Convex and Conical Spaces}
\author{Reynald Affeldt\orcidID{0000-0002-2327-953X}\inst{1} \and
  Jacques Garrigue\orcidID{0000-0001-8056-5519}\inst{2} \and
  Takafumi Saikawa\orcidID{0000-0003-4492-745X}\inst{2} }
\institute{National Institute of Advanced Industrial Science and Technology, Japan
  \and
  Nagoya University, Japan}

\def\newterm#1{{\sl #1}}
\def\infotheo{\textsc{InfoTheo}}
\def\coq{\textsc{Coq}}
\def\mathcomp{\textsc{MathComp}}
\def\coqin#1{\mintinline{ssr}{#1}}
\newcommand\pchoice[3]{#1 \triangleleft {_{#2}} \triangleright #3}
\newcommand\pchoiceS[1]{\triangleleft {_{#1}} \triangleright}
\newcommand\enum[1]{\{ #1 \}}
\def\convnsymbol{{\lhd\mspace{-3.5mu}\rhd}}
\DeclareMathOperator*{\narypchoiceS}{\convnsymbol}
\newcommand\narypchoice[4]{\narypchoiceS_{#3<#4} {#1} {#2}}
\newcommand\comprh[2]{\left\{{#1}\ \middle|\ {#2}\right\}}
\newcommand\imp{\Rightarrow}

\newcommand{\citefile}[2]{\cite[{\tt\small {#1}}]{#2}}

\newcommand\Sone[1]{\iota(#1)}
\let\SoneS\iota
\def\ska#1#2{{#1 * #2}}

\begin{document}
\maketitle

\def\rei#1{{\color{red} [NB(rei): #1]}}
\def\jac#1{{\color{blue} [NB(jacques): #1]}}
\def\sai#1{{\color{green} [NB(saikawa): #1]}}

\begin{abstract}
Convex sets appear in various mathematical theories, and are
used to define notions such as convex functions and hulls.
As an abstraction from the usual definition of convex sets in vector spaces, we
formalize in Coq an intrinsic axiomatization of convex sets, namely convex
spaces, based on an operation taking barycenters of points.
A convex space corresponds to a specific type that does not refer to a
surrounding vector space.  This simplifies the definitions of functions on it.
We show applications including the convexity of information-theoretic functions
defined over types of distributions.
We also show how convex spaces are embedded in conical spaces, which are
abstract real cones, and use the embedding as an effective device to ease
calculations.
\end{abstract}

\section{Introduction}

The notion of convex sets appears in various mathematical theories.  A subset
$X$ of a real vector space is called a convex set if, for any $x, y\in X$ and
$p\in [0,1]$, their \newterm{convex combination} $px+(1-p)y$ is again in $X$.
One basic use of it is to define the convexity of functions.  A function $f$ is
said to be convex if $f(px+(1-p)y) \leq pf(x)+(1-p)f(y)$ for any convex
combination $px+(1-p)y$. Thus, convex sets are natural domains for convex
functions to be defined on.
Good examples of these notions can be found in information theory,
where convexity is a fundamental property of important functions such
as logarithm, entropy, and mutual information.  Our \infotheo{}
library~\cite{infotheo} developed in the \coq{} proof
assistant~\cite{coq} has a formalization of textbook
proofs~\cite{cover2006} of such results.

In the course of formalizing such convexity results, we find that axiomatizing
convex sets is a useful step which provides clarity and organizability in the
results.  We abstract the usual treatment of convex sets as subsets of some
vector space and employ an algebraic theory of \newterm{convex spaces}, which was
introduced by Stone~\cite{stone1949annali}.  The formalization uses the \newterm{packed
class} construction \cite{garillot2009tphols,mahboubi2013itp}, so as to obtain generic
notations and lemmas, and more importantly, to be able to combine structures.
Binary convex spaces are formalized in Sect.~\ref{sect:convex_space},
and their multiary versions are formalized in Sect.~\ref{sect:convn},
along with proofs of equivalence.

We also formalize an embedding of convex spaces into \newterm{conical spaces}
(a.k.a.\ cones or real cones~\cite{varacca2006mscs}), which we find an
indispensable tool to formalize convex spaces.
Examples in the literature avoid proving properties of convex spaces directly
and choose to work in conical spaces.  This is especially the case when their
goal can be achieved either way~\cite{kirch1993master, varacca2006mscs}.  Some
authors suggest 
that the results in conical spaces can be backported to
convex spaces~\cite{flood1981jams, keimel2016lmcs}.  We apply this
method in Sect.~\ref{sect:conical} to
enable additive handling of convex combinations.
By formalizing the relationship between convex and conical spaces, we work out
short proofs of a number of lemmas on convex spaces.  Among them is Stone's key
lemma~\cite[Lemma~2]{stone1949annali}, whose proof is often omitted in the
literature despite its fundamental role in the study of convex spaces.

We complete this presentation with applications of our formalization
to convex hulls (Sect.~\ref{sect:convex_hulls}) and to convex
functions (Sect.~\ref{sect:convex_functions}).

While our proofs do not introduce extra axioms, some libraries used in
our development, such as mathcomp-analysis~\cite{cohen2018jfr},
contain axioms which make parts of our work classical. In
particular, our definition of convex sets is based on classical sets,
assuming decidable membership.

\section{Convex spaces}
\label{sect:convex_space}

Let us begin with the definition of convex spaces.  As mentioned in the
introduction, convex spaces are an axiomatization of the usual notion of convex
sets in vector spaces.  It has a long history of repeated reintroduction by many
authors, often with minor differences and different names: barycentric
algebra~\cite{stone1949annali}, semiconvex algebra~\cite{swirszcz1974bapmam},
or, just, convex sets~\cite{jacobs2010tcs}.

We define convex spaces following Fritz~\cite[Definition 3.1]{fritz2015arxiv}.
\begin{definition}[
  \coqin{Module ConvexSpace} in \cite{convexchoice}]
\label{def:convex_space}
A convex space is a structure for the following signature:
\begin{itemize}
\item Carrier set $X$.
\item Convex combination operations $(\pchoice \_ p \_) : X \times X \to X$
  indexed by $p\in [0,1]$.
\item Unit law: $\pchoice x 1 y = x$.
\item Idempotence law: $\pchoice x p x = x$.
\item Skewed commutativity law: $\pchoice x {1-p} y = \pchoice y p x$.
\item Quasi-associativity law: $\pchoice x p {({\pchoice y q z})} =
  \pchoice {(\pchoice x r y)} s z$, \\
  where $s=1-(1-p)(1-q)$ and
  $r=
  \begin{cases}
    p/s & \text{if $s\not= 0$} \\
    0 & \text{otherwise}
  \end{cases}$. \\
  (Note that $r$ is irrelevant to the value of
  $\pchoice {(\pchoice x r y)} s z$ if $s=0$.)
\end{itemize}
\end{definition}

We can translate this definition to \coq{} as a \newterm{packed
  class}~\cite{garillot2009tphols} with the following mixin interface:
\begin{minted}[numbers=left,escapeinside=77]{ssr}
Record mixin_of (T : choiceType) : Type := Mixin {
  conv : prob -> T -> T -> T where "a <| p |> b" := (conv p a b);
  _ : forall a b, a <| 1%:pr |> b = a ; 7\label{line:prop1}7
  _ : forall p a, a <| p |> a = a ;
  _ : forall p a b, a <| p |> b = b <| p.~%:pr |> a;
  _ : forall (p q : prob) (a b c : T),
      a <|7\,7p7\,7|> (b <|7\,7q7\,7|> c) = (a <|[r_of p, q]|> b) <|7\,7[s_of p, q]7\,7|> c }.
\end{minted}
There are some notations and definitions to be explained.  The type \coqin{prob} in
the above \coq{} code denotes the closed unit interval $[0,1]$.  The notation \mintinline{ssr}{r
is a notation for a real number \coqin{r} equipped with a canonical proof that
$0 \leq \coqin{r} \leq 1$. The notation \coqin{p.~} is for $1 -
\coqin{p}$. The notation \coqin{[s_of p, q]} is for $1-(1-\coqin{p})(1-\coqin{q})$,
and \coqin{[r_of p, q]} for $\coqin{p} / \coqin{[s_of p, q]}$.

Intuitively, one can regard the convex combination as a probabilistic choice
between two points.  At line~\ref{line:prop1}, the left argument is chosen with
probability~$1$.
The lines that follow correspond to idempotence, skewed commutativity,
and quasi-associativity.

An easy example of convex space is the real line $\mathbb R$, whose
convex combination is expressed by ordinary addition and
multiplication as $pa + (1 - p)b$.  Probability distributions also
form a convex space.  In the formalization, the type \coqin{fdist A}
of distributions over any finite type \coqin{A}~(borrowed from
previous work~\cite{affeldt2014jar}) is equipped with a convex space
structure, where the convex combination of two distributions
$d_1, d_2$ is defined pointwise as
$x \mapsto pd_1(x) + (1 - p)d_2(x)$.

As a result of the packed class construction, we obtain the type
\coqin{convType} of all types which implicitly carry the above axioms.  Then,
each example of convex space is declared to be canonically a member of
\coqin{convType}, enabling the implicit inference of the appropriate convex
space structure.  These two implicit inference mechanisms combined make the
statement of generic lemmas on convex spaces simple and applications easy.

\section{Multiary convex combination}
\label{sect:convn}

Convex spaces can also be characterized by multiary convex combination
operations, which combine finitely many points $x_0, \dots, x_{n-1}$ at once,
according to some finite probability distribution $d$ over the set
$I_n = \enum{0, \dots, n-1}$, i.e., $d_i \geq 0$ and $\sum_{i<n} d_i = 1$.
In this section we consider different axiomatizations, and their
equivalence with the binary axioms.

\subsection{Axiomatization}
A definition of convex spaces based on multiary operations is given as follows
(see for example~\cite[Definition 5]{bonchi2017concur} and
\cite[Sect.~2.1]{heerdt2018ictac}).
\begin{definition}[Convex space, multiary version]
\label{def:convex_space_multiary}
A convex space based on multiary operations is a structure for the following
signature:
\begin{itemize}
\item Carrier set $X$.
\item Multiary convex combination operations,
  indexed by an arity $n$ and a distribution $d$ over $I_n$:
  \[
    \begin{array}{ccc}
      X^n & \to & X \\
      (x_i)_{i<n} & \mapsto & \displaystyle\narypchoiceS_{i<n} d_i x_i
    \end{array}
  \]
\item Projection law: if $d_j = 1$, 
\(\displaystyle\narypchoice{d_i}{x_i}in=x_j\).
\hfill (\coqin{ax_proj} in \cite{convexchoice})
\item Barycenter law:
\(\displaystyle
\narypchoice{d_i}{\left(\narypchoice{e_{i,j}}{x_j}{j}{m}\right)}{i}{n} =
  \narypchoice{\left(\sum_{i<n}d_ie_{i,j}\right)}{x_j}{j}{m}\).
\hfill (\coqin{ax_bary} in \cite{convexchoice})
\end{itemize}
\end{definition}
Note that in our \coq{} code, $\narypchoice{d_i}{x_i}{i}{n}$
appears as \mintinline{ssr}{<&>_d x} or \coqin{altConvn d x},
indicating more
explicitly that the operation takes two arguments $d$ and $x$.

This multiary convex structure and the binary one given in
Sect.~\ref{sect:convex_space} are equivalent: the
multiary and binary operators interpret each other satisfying the needed axioms,
and the interpretations cancel out when composed.
While the binary axiomatization is easy to instantiate, the multiary
version exhibits the relationship to probability distributions.
Therefore we want to establish this equivalence before working further on other
constructions over convex spaces.

In the literature, this equivalence is justified without much detail by referring to the
seminal article by Stone~\cite{stone1949annali} (see, e.g., \cite[Theorem
4]{jacobs2010tcs}, \cite[Proposition 7]{bonchi2017concur}).
Yet, what Stone gave is not an explicit axiomatization of the multiary
convex operator, but a number of lemmas targeted at proving an
embedding of (binary) convex spaces into vector spaces. These lemmas
include the following one, that is seen as a justification for the
barycenter law in the binary axiomatization.
\begin{lemma}[Lemma 4 in \cite{stone1949annali}]
If the given masses and their associated points are partitioned into
groups (of non-zero total masses) in any way, then the center of mass
is identical with that of masses equal to the respective total masses
for the various groups, each placed at the center of mass for
the corresponding group.
\end{lemma}
The relation to the barycenter law is implied if one sees a convex combination
$\narypchoice{(\sum_{i<n}d_ie_{i,j})}{x_j}{j}{m}$ as a point defined in terms of
a set of generating points $\{x_j\}_{j<m}$ (they generate their convex hull).
Then $\narypchoice{d_i}{\left(\narypchoice{e_{i,j}}{x_j}{j}{m}\right)}{i}{n}$
corresponds to grouping the generating points by filtering through the
distributions $\{e_i\}_{i<n}$.
But this grouping is not necessarily a partition since there could be
shared elements, hence the relation is not direct.

Beaulieu~\cite[Def. 3.1.4]{beaulieu2008phd} proposed an
alternative multiary axiomatization, which was actually presented
as a model for countable probabilistic choice (rather than a definition
of convex space). His partition law
corresponds exactly to the statement of Stone's lemma.
\begin{definition}[Convex space, Beaulieu style]
A convex space is a structure for the previous operations
$\narypchoice{d_i}{}{i}{n}$ and the following laws.
\begin{itemize}\itemsep=1.5ex
\item Partition law: ~
\(\displaystyle \narypchoiceS_{i\in I}\lambda_i x_i =
\narypchoiceS_{j\in J} \rho_j\left(\narypchoiceS_{k\in
    K_j} \frac{\lambda_k}{\rho_j}x_k\right)\)
\hfill (\coqin{ax_part} in \cite{convexchoice}) \\
where $\{K_j\mid j\in J\}$ is a partition of $I$, and $\rho_j =
\sum_{k\in K_j}\lambda_k\neq 0$.
\item Idempotence law: ~
\(\displaystyle \narypchoiceS_{i\in I} \lambda_i A_i = A\) ~
if $A_i = A$ for all $\lambda_i > 0$.
\hfill (\coqin{ax_idem} in \cite{convexchoice})
\end{itemize}
\end{definition}
In the implementation, using sets as indexing domains of the combination
operators would be cumbersome, so that the partition law is actually
expressed as follows, using a map $\check{K}$ and Kronecker's $\delta$.
\[ \narypchoiceS_{i < n}\lambda_i x_i =
\narypchoiceS_{j < m} \rho_j\left(\narypchoiceS_{k < n}
\delta_{j,\check{K}(k)}\frac{\lambda_k}{\rho_j}x_k\right)
\quad \mbox{where } \check{K} : I_n \to I_m, ~ K_j = \check{K}^{-1}(j)
\]
We also have to separately show that
$(\delta_{j,\check{K}(k)}\frac{\lambda_k}{\rho_j})_{k<n}$ and $(\rho_j)_{j<m}$
form probability distributions.
As an exceptional case,
$(\delta_{j,\check{K}(k)}\frac{\lambda_k}{\rho_j})_{k<n}$ is replaced by a
uniform distribution if $\rho_j = 0$.

\subsection{Equivalence of axiomatizations}

After considering the different axiomatizations, we decided to prove a
triangular equivalence: between multiary convex structures in
standard and Beaulieu style, and then with the binary convex structure
given in Sect.~\ref{sect:convex_space}.
The relations we will explain in this section are depicted in Fig.~\ref{fig:relations}.

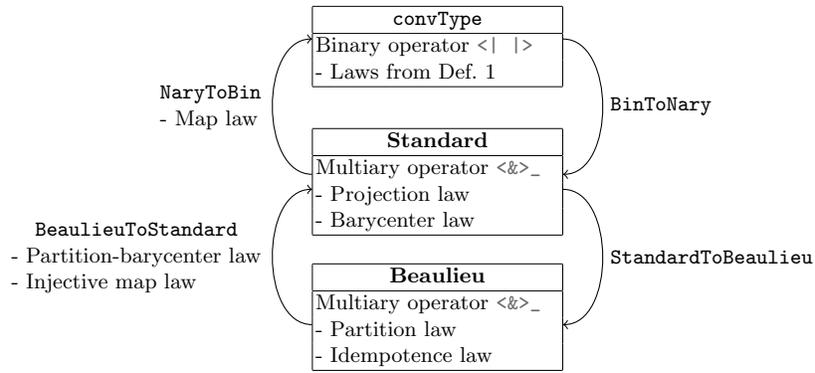
\begin{figure}
\centering
\begin{tikzpicture}[
  myarrow/.style={->},
  scale=0.9,
  every node/.style={scale=0.9}
  ]

\node (beaulieu) {\begin{tabular}{|p{3.6cm}|}
                    \hline
                    \multicolumn{1}{|c|}{{\bf Beaulieu}} \\
                    \hline
                    Multiary operator \coqin{<&>_} \\
                    - Partition law \\
                    - Idempotence law \\
                    \hline
                  \end{tabular}};
\node (standard)[above of=beaulieu,yshift=1cm] {\begin{tabular}{|p{3.6cm}|}
                                       \hline
                                       \multicolumn{1}{|c|}{{\bf Standard}} \\
                                                  \hline
                                       Multiary operator \coqin{<&>_} \\
                                       - Projection law \\ 
                                       - Barycenter law \\
                                       \hline
                                     \end{tabular}};
\path (beaulieu) edge[myarrow,out=180,in=180,transform canvas={yshift=-1mm,xshift=1mm}] node[left] {\begin{tabular}{l}
  \multicolumn{1}{c}{\coqin{BeaulieuToStandard}} \\
  - Partition-barycenter law \\
  - Injective map law \end{tabular}} (standard);
\path (standard) edge[myarrow,out=0,in=0,transform canvas={yshift=-1mm,xshift=-1mm}] node[right] {\coqin{StandardToBeaulieu}} (beaulieu);
\node (convtype)[above of=standard,yshift=1cm] {
  \begin{tabular}{|p{3.6cm}|}
    \hline
    \multicolumn{1}{|c|}{\coqin{convType}} \\
    \hline
    Binary operator \coqin{<|  |>}  \\
    - Laws from Def.~\ref{def:convex_space} \\
    \hline
  \end{tabular}};
\path (standard) edge[myarrow,out=180,in=180,transform canvas={yshift=1mm,xshift=1mm}] node[left] {\begin{tabular}{l} \multicolumn{1}{c}{\coqin{NaryToBin}}
 \\ - Map law\end{tabular}} (convtype);
\path (convtype) edge[myarrow,out=0,in=0,transform canvas={yshift=1mm,xshift=-1mm}] node[right] {\coqin{BinToNary}} (standard);
\end{tikzpicture}
\caption{Relations between the various formalizations of convex spaces}
\label{fig:relations}
\end{figure}

The first equivalence, between multiary convex axioms, is rather
technical. The first direction, proving Beaulieu's axioms from the
standard presentation (functor \coqin{StandardToBeaulieu}
in~\cite{convexchoice}), is relatively easy, as the partition law is
intuitively just a special case of the barycenter law, where supports%
\footnote{The support of a probability distribution $d$ is the set
\(\{i \mid d_i > 0\}\).}
are disjoint, and the idempotence
law can be derived as a combination of
the two standard laws. However, the second direction (functor
\coqin{BeaulieuToStandard}) is harder, and
led us to introduce two derived laws:
\begin{itemize}\it
\item Partition-barycenter law: barycenter law, with disjoint supports.
\hfill (\coqin{ax_bary_part})
\item Injective map law: \(\displaystyle
  \narypchoice{d_i}{g_{u(i)}}{i}{m} =
  \narypchoice{\sum_{\substack{i < m \\ u(i)=j}}d_i}{g_j}{j}{n}\)
  with $u$ injective.
  \hfill (\coqin{ax_inj_map})
\end{itemize}
The partition-barycenter law can be derived from the Beaulieu style
axioms, and in turn is used to prove the injective map law.
Together they allow to prove the barycenter law.

The equivalence between binary and multiary axiomatizations requires first
to define their operators in terms of each other.
\begin{definition}[
  \coqin{Convn} and \coqin{binconv} in \cite{convexchoice}]
\label{def:Convn}
\begin{enumerate}[(a)]
\item
  Let $d : I_n \to [0,1]$ be a finite distribution,
  and $x : I_n \to X$ be points in a convex space~$X$.
  Then the multiary convex combination of these points and distribution is
  defined from the binary operator by recursion on~$n$ as follows:
  \begin{align*}
    \narypchoice{d_i}{x_i}{i}{n}
    &=
      \begin{cases}
        x_0 & \text{if $d_0 = 1$ or $n = 1$}\\
        \displaystyle
        \pchoice{x_0}{d_0}{\left(
          \narypchoice{d'_i}{x_{i+1}}{i}{n-1}
          \right)}
        & \text{otherwise}
      \end{cases} \\
    &\text{where $d'$ is a new distribution: $d'_i = d_{i+1} / (1 - d_0)$ .}
  \end{align*}
\item
  Let $p$ be a probability and $x_0,x_1$ be points in a convex space.
  Then their binary combination is defined from the multiary operator as
  follows:
\[ \pchoice{x_0}{p}{x_1} = \narypchoice{d_i}{x_i}{i}{2}
\quad\quad \mbox{where } d_0 = p \mbox{ and } d_1 = 1-p. \]
\end{enumerate}
\end{definition}

The first direction, functor \coqin{BinToNary} in \cite{convexchoice},
must prove that the
first definition satisfies the multiary axioms, and indeed amounts to
proving a variant of Stone's lemma. We will see in the next section
that the original proof by Stone is better formalized by
transporting the argument to conical spaces.

The opposite direction, functor \coqin{NaryToBin}, must prove the binary
axioms from the multiary ones.
While we start from the standard version, the
idempotence law proved to be instrumental in this task, together with
the following unrestricted map law.
\begin{itemize}\it
\item Map law: \(\displaystyle
  \narypchoice{d_i}{g_{u(i)}}{i}{m} =
  \narypchoice{\sum_{\substack{i < m \\ u(i)=j}}d_i}{g_j}{j}{n}\)
for any map $u$.
\hfill (\coqin{ax_map} in \cite{convexchoice})
\end{itemize}

Finally, one also needs to prove that the definitions we used for each
operation in both directions are coherent.
\begin{lemma}[\coqin{equiv_conv} and \coqin{equiv_convn} in
    \cite{convexchoice}]
The constructions in Def.~\ref{def:Convn} (\coqin{Convn} and \coqin{binconv}) cancel each other. That is,
\begin{itemize}
\def\fst{^*}
\def\snd{^\dagger}
\item If $\narypchoiceS\fst$ is the operator induced by Def.~\ref{def:Convn}(a),
and $\pchoiceS{\_}\snd$ the one induced from it by Def.~\ref{def:Convn}(b),
we can derive \( a \pchoiceS{p}\snd b = \pchoice{a}{p}{b} \) from the
binary axioms.
\item If $\pchoiceS{\_}\fst$ the operator induced by Def.~\ref{def:Convn}(b),
$\narypchoiceS\snd$ is the one induced from it by Def.~\ref{def:Convn}(a),
we can derive \(\narypchoiceS\snd_{i<n}d_ix_i = \narypchoice{d_i}{x_i}{i}{n}\)
from the multiary axioms.
\end{itemize}
\end{lemma}

\section{Conical spaces and embedded convex spaces}
\label{sect:conical}

The definition of multiary convex combination operator in the previous section
(Def.~\ref{def:Convn}(a)) relied on recursion.
This makes the definition look complicated, and moreover, the algebraic
properties of the combination difficult to see.
If we consider the special case of convex sets in a vector space, the meaning of
multiary combinations and the algebraic properties become evident:
\[
  \narypchoice{d_i}{x_i}{i}{n}
  = d_0 x_0 + \dots + d_{n-1} x_{n-1}.
\]
The additions on the right-hand side are of vectors, and thus are associative
and commutative.
This means that the multiary combination on the left-hand side is invariant
under permutations or partitions on indices.
We want to show that these invariance properties are also satisfied generally in
any convex space.

However, the search for the proofs is painful if naively done.  This is because
binary convex combination operations satisfy associativity and commutativity
only through cumbersome parameter computations.
For example, a direct proof of the permutation case involves manipulations 
on the set $I_n$ of indices and on the symmetry groups,
which require fairly long combinatorics~\cite[Lemma~2]{stone1949annali}.

We present a solution to this complexity by transporting the arguments on convex
spaces to a closely related construction of conical spaces.  Conical spaces are
an abstraction of cones in real vector spaces just like convex spaces are an
abstraction of convex sets.  Like convex spaces, the definition of conical
spaces appears in many articles.  We refer to the ones by Flood (called semicone
there)~\cite{flood1981jams} and by Varacca and Winskel (called real cone
there)~\cite{varacca2006mscs}:
\begin{definition}[Conical space]
\label{def:conical_space}
A conical space is a semimodule over the semiring of non-negative reals.
That is, it is a structure for the following signature:
\begin{itemize}
\item Carrier set $X$.
\item Zero $\mathbf 0 : X$. 
\item Addition operation $\_ + \_ : X \times X \to X$.
\item Scaling operations $c \_ : X \to X$ indexed by $c \in \mathbb R_{\geq 0}$.
\item Associativity law for addition: $x + (y + z) = (x + y) + z$.
\item Commutativity law for addition: $x + y = y + x$.
\item Associativity law for scaling: $c (d x) = (c d) x$.
\item Left-distributivity law: $(c + d)x = cx + dx$.
\item Right-distributivity law: $c(x + y) = cx + cy$.
\item Zero law for addition: $\mathbf 0 + x = x$.
\item Left zero law for scaling: $0 x = \mathbf 0$.
\item Right zero law for scaling: $c \mathbf 0 = \mathbf 0$.
\item One law for scaling: $1 x = x$.
\end{itemize}
\end{definition}

We display this definition only to show that conical spaces have
straightforward associativity and commutativity. In fact, the formalization is
elaborated on the embedding of convex spaces into canonically constructed
conical spaces, which appeared in the article by Flood~\cite{flood1981jams}.
We build on top of each convex space $X$, the conical space $S_X$ of its
``scaled points'':
\begin{definition}[
  \coqin{scaled_pt}, \coqin{addpt}, and \coqin{scalept} in \cite{convexchoice}]
\label{def:conical_space_scaled_pt}
  Let $X$ be a convex space.  We define a set $S_X$ which becomes a conical
  space with the following addition and scaling operations.
  \[
    S_X := (\mathbb R_{>0} \times X) \cup \{\mathbf 0\}.
  \]
  That is, the points of $S_X$ are either a pair $\ska{p}{x}$ of
  $p\in\mathbb R_{>0}$ and $x\in X$, or a new additive unit $\mathbf 0$.
  Addition of points $a, b \in S_X$ is defined by cases to deal with
  $\mathbf 0$:
  \[
    a + b :=
    \begin{cases}
      \ska{(r+q)}{(\pchoice{x}{r/(r+q)}{y})} & \text{if $a=\ska{r}{x}$ and $b=\ska{q}{y}$}\\
      a & \text{if $b = \mathbf 0$}\\
      b & \text{if $a = \mathbf 0$}
    \end{cases}
  \]
  Scaling $a \in S_X$ by $p \in \mathbb R_{\geq 0}$ is also defined by cases:
  \[
    pa :=
    \begin{cases}
      \ska{pq}{x} & \text{if $p > 0$ and $a = \ska{q}{x}$} \\
      \mathbf 0 & \text{otherwise}
    \end{cases}
  \]
\end{definition}
We omit here the proofs that $S_X$ with these addition and scaling satisfies the
conical laws. They are proved formally in \cite{convexchoice} (see the
lemmas \coqin{addptC}, \coqin{addptA}, \coqin{scalept_addpt}, etc.).

Properties of the underlying convex spaces are transported
into and back from this conical space, through an embedding:
\begin{definition}[\coqin{S1} in \cite{convexchoice}]
\[
  \SoneS :
  \begin{array}[t]{ccc}
  X & \rightarrowtail & S_X\\
  x & \mapsto & \ska{1}{x}
\end{array}
\]
\end{definition}

Convex combinations in $X$ are mapped by $\SoneS$ to additions in $S_X$.
\begin{lemma}[\coqin{S1_convn} in \cite{convexchoice}]
   \label{lem:S1_convn}
   \[
     \Sone{\narypchoice{d_i}{x_i}{i}{n}} = \sum_{i<n} d_i \Sone{x_i}.
   \]
\end{lemma}
The right-hand side of the lemma is a conical sum (Fig.~\ref{fig:ex_S1_convn}), which
behaves like an ordinary linear sum thanks to the conical laws, and enjoys good
support from \mathcomp{}'s big operator library~\cite{bertot2008tphols}.

\begin{figure}[H]
  \centering
  \begin{tabular}[t]{ccc}
    \raisebox{1.5cm}{\includegraphics[scale=0.4]{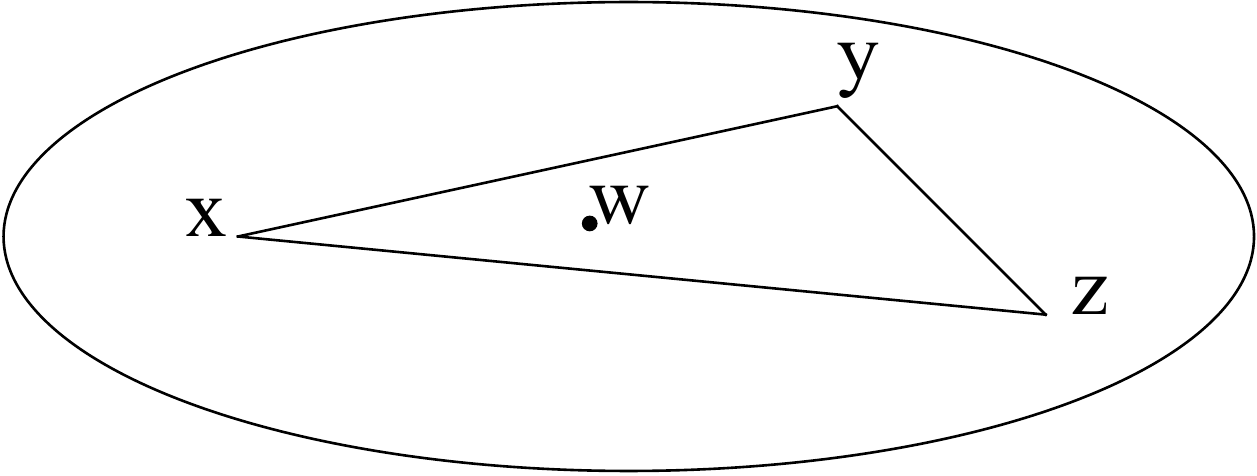}}
    &
      \raisebox{2.3cm}{\quad $\stackrel \iota \longrightarrow$ \quad }
    &
      \includegraphics[scale=0.4]{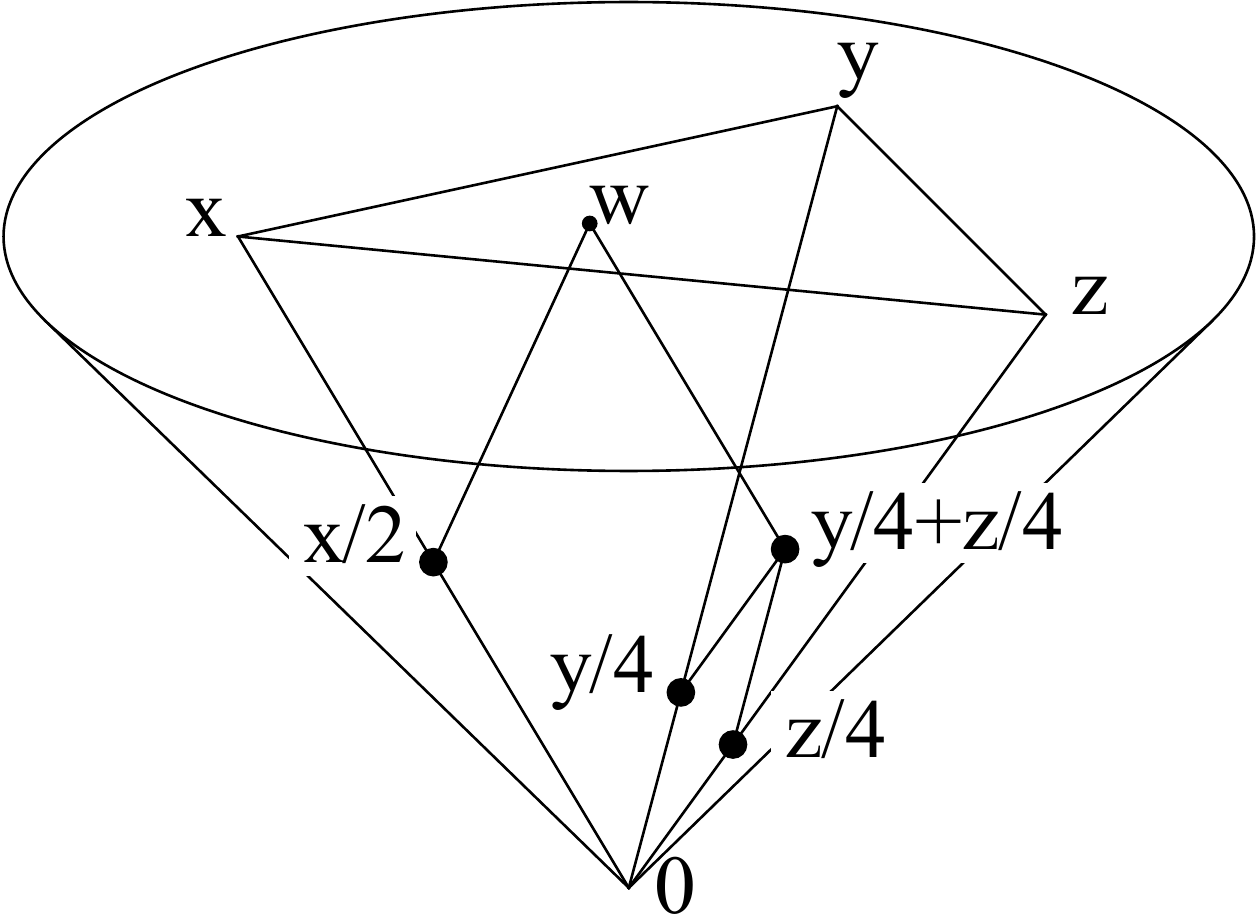}
  \end{tabular}
  \caption{Example of \coqin{S1_convn}:
    $\ska{1}{w} = \ska{\frac 1 2}x + \ska{\frac 1 4}y + \ska{\frac 1 4}z$}
  \label{fig:ex_S1_convn}
\end{figure}

With these preparations, properties such as~\cite[Lemma 2]{stone1949annali}
can be proved in a few lines of \coq{} code:
\begin{lemma}[\coqin{Convn_perm} in \cite{convexchoice}]
  \[
    \narypchoice{d_i}{x_i}{i}{n} =
    \narypchoice{(d\circ s)_i}{(x\circ s)_i}{i}{n},
  \]
  where $s$ is any permutation on the set of indices $n$.
\end{lemma}

The proof of the barycenter property~\cite[Lemma~4]{stone1949annali} from
Sect.~\ref{sect:convn} is based on the same technique (see
\coqin{Convn_convnfdist} in \cite{convexchoice}).

A way to understand this conical approach is to start from Stone's definition of
convex spaces~\cite{stone1949annali}.  He uses a quaternary convex operator
$(x,y;\alpha,\beta)$ where $x$ and $y$ are points of the space, and $\alpha$ and
$\beta$ are non-negative coefficients such that $\alpha+\beta>0$.
Its values are quotiented by an axiom to be invariant under
scaling, removing the need to normalize coefficients for associativity.  This
amounts to regarding a convex space as the projective space of some conical
space.

The definition of $S_X$ is a concrete reconstruction of such a conical space
from a given convex space $X$.  The benefit of this method over Stone's is the
removal of quotients by moving the coefficients from operations to values.
We can then use the linear-algebraic properties of conical sums such as the
neutrality of zeroes, which had to be specially handled in Stone's
proofs (e.g., \cite[Lemma 2]{stone1949annali}).

\paragraph{Examples}
We illustrate how $\SoneS$ is used in practice with the proof of the
\newterm{entropic identity}.  Let \coqin{T} be a \coqin{convType}; we
want to show that
\begin{align}
\pchoice{(\pchoice{a}{q}{b})}{p}{(\pchoice{c}{q}{d})} =
\pchoice{(\pchoice{a}{p}{c})}{q}{(\pchoice{b}{p}{d})}. \label{eqn:convACA}
\end{align}
We could use the properties of convex spaces, but this will result
in cumbersome computations, in particular because of
quasi-associativity.
Instead, we proceed by an embedding into the set of scaled points over
\coqin{T} using $\SoneS$.
First, we observe that these scaled points form a convex space
for the operator $p, a, b \mapsto pa + \bar{p}b$ and that
$\Sone{\pchoice{a}{p}{b}} = \pchoice{\Sone a}{p}{\Sone b}$.
As a consequence, when we apply $\SoneS$ to Equation~(\ref{eqn:convACA}),
its left-hand side becomes
$$
p(q \Sone a + \bar{q}\Sone b) +
\bar{p}(q\Sone c + \bar{q}\Sone d).
$$
The main difference with Equation~(\ref{eqn:convACA}) is that
$+$ (\coq{} notation: \coqin{addpt}) enjoys (unconditional) associativity, making
the rest of the proof easier.
In the proof script below, line~\ref{line:embed} performs the
embedding by first using the injectivity of $\SoneS$ (lemma
\coqin{S1_inj}), then using the fact that $\SoneS$ is a morphism w.r.t.\
$\pchoice{\_}{p}{\_}$ (lemma \coqin{S1_conv}), and last by
revealing the definition of the operator of the convex spaces formed
by scaled points (lemma \coqin{convptE}).
The proof can be completed by rewritings with properties of
\coqin{addpt} and \coqin{scalept} until the left-hand side matches the
right-hand side.

\begin{minted}[numbers=left,escapeinside=77]{ssr}
Lemma convACA (a b c d : T) p q :
  (a <|q|> b) <|p|> (c <|q|> d) = (a <|p|> c) <|q|> (b <|p|> d).
Proof.
apply S1_inj; rewrite ![in LHS]S1_conv !convptE. 7\label{line:embed}7
rewrite !scalept_addpt ?scalept_comp //.
rewrite !(mulRC p) !(mulRC p.~) addptA addptC (addptC (scalept (q * p) _)).
rewrite !addptA -addptA -!scalept_comp -?scalept_addpt //.
by rewrite !(addptC (scalept _.~ _)) !S1_conv.
Qed.
\end{minted}

\smallskip

We conclude this section with an example that provides a closed
formula for the multiary convex combination
$\narypchoiceS_{i<n} e_i g_i$ (\coq{} notation:
\coqin{<|>_e g}) in the case of the real line (seen as a convex
space):
\begin{minted}[numbers=left,escapeinside=77]{ssr}
Definition scaleR x : R := if x is p *: y then p * y else 0.
Definition big_scaleR := big_morph scaleR scaleR_addpt scaleR0.
Lemma avgnRE n (g : 'I_n -> R) e : <|>_e g = \sum_(i < n) e i * g i.
Proof.
rewrite -[LHS]Scaled1RK S1_convn big_scaleR.
by under eq_bigr do rewrite scaleR_scalept // Scaled1RK.
Qed.
\end{minted}
This corresponds to the following transformations of the left-hand
side.
\[ \begin{array}{rl@{\mbox{~~~ by }}l}
\narypchoice{e_i}{g_i}{i}{n} \, = &
\coqin{scaleR}(\Sone{\narypchoice{e_i}{g_i}{i}{n}}) &
\coqin{Scaled1RK} \\
= & \coqin{scaleR}(\sum_{i<n} e_i\Sone{g_i}) & \coqin{S1_convn} \\
= & \sum_{i<n} \coqin{scaleR}(e_i\Sone{g_i}) & \coqin{big_scaleR} \\
= & \sum_{i<n} e_i\coqin{scaleR}(\Sone{g_i}) & \coqin{scaleR_scalept} \\
= & \sum_{i<n} e_i g_i & \coqin{Scaled1RK}
\end{array} \]

\section{Formalization of convex sets and hulls}
\label{sect:convex_hulls}

\def\hull#1{{\textsf{hull}\left(#1\right)}}

Our first application of convex and conical spaces is the
formalization of convex sets and convex hulls.  Besides mathematics,
they also appear in many applications of convex spaces such as program
semantics~\cite{beaulieu2008phd,cheung2017phd}.

\begin{definition}
[\coqin{is_convex_set} in \cite{convexchoice}]\label{def:convex_subset}
Let $T$ be a convex space.
A subset $D$ in~$T$ is a convex set if, for any $p\in[0,1]$
and $x, y\in D$, $\pchoice x p y \in D$.
\end{definition}
We use the predicate \coqin{is_convex_set} to define
the type \coqin{{convex_set T}} of convex sets over~\coqin{T}.

We can turn any set of points in a convex space into a convex set,
namely, by taking \newterm{convex hulls}.

\begin{definition}[\coqin{hull} in \cite{convexchoice}]
  For a subset $X$ of $T$, its hull $\overline X$ is
  \[
    \overline X = \comprh{\narypchoice{d_i}{x_i}{i}{n}}
    { n\in \mathbb N \land
      d \mbox{ is a distribution over } I_n \land
      \forall i < n,\ x_i\in X
    }.
  \]
\end{definition}

\paragraph{Example}
The following example illustrates the usefulness of conical spaces
when reasoning about convex hulls.

Our goal is to prove that for any $z \in \hull{X\cup Y}$
($X\neq\emptyset$, $Y\neq\emptyset$), there exist $x \in X$ and
$y \in Y$ such that $z = \pchoice x p y$ for some $p$ (see the formal
statement at line \ref{line:hullstatement} below).

We first introduce two notations.
Let \coqin{scaled_set X} be the set
$\{ \ska{p}{x}\,|\,x\in \texttt{X}\}$.
For any $a\neq 0$, let \coqin{[point of a0]} (where \coqin{a0} is the
proof that $a\neq 0$) be the $x$ such that $a=\ska{p}{x}$ for some~$p$.

To prove our goal, it is sufficient to prove that there exist
$a \in \coqin{scaled_set X}$ and $b \in \coqin{scaled_set Y}$ such that
$\Sone{z} = a + b$ (this reasoning step is the purpose of line \ref{line:hullsuff}).
When $a=0$ or $b=0$, we omit easy proofs at
lines~\ref{line:hullx0} and \ref{line:hully0}.
Otherwise, we can take $x$ to be \coqin{[point of a0]} and $y$ to be
\coqin{[point of b0]} as performed by the four lines from line~\ref{line:no0}.

We now establish the sufficient condition (from line \ref{line:hullmain}).
Since $z$ is in the hull, we have a distribution $d$ and $n$ points
$g_i$ such that $z = \narypchoice{d_i}{g_i}{i}{n}$. We then decompose
$\Sone{z}$ as follows:
$$\Sone{z} = \sum_{i < n} d_i(\Sone{g_i}) = \underbrace{\sum_{i < n,g_i\in X} d_i(\Sone{g_i})}_b + \underbrace{\sum_{i < n, g_i\notin X} d_i(\Sone{g_i})}_c.$$
We conclude by observing that $b$ is in \coqin{scaled_set X}
and that $c$ is in \coqin{scaled_set Y} because
$\{g_i|g_i\notin X\} \subseteq Y$.

\begin{minted}[numbers=left,escapeinside=77]{ssr}
Lemma hull_setU (z : T) (X Y : {convex_set T}) : X !=set0 -> Y !=set0 -> 7\label{line:hullstatement}7
  hull (X `|` Y) z ->
  exists2 x, x \in X & exists2 y, y \in Y & exists p, z = x <| p |> y.
Proof.
move=> [dx ?] [dy ?] [n -[g [d [gT zg]]]].
suff [a] : exists2 a, a \in scaled_set X & exists2 b, b \in scaled_set Y & 7\label{line:hullsuff}7
    S1 z = addpt a b.
  have [/eqP -> _ [b bY]|a0 aX [b]] := boolP (a == Zero) by ... 7\label{line:hullx0}7
  have [/eqP -> _|b0 bY] := boolP (b == Zero) by ... 7\label{line:hully0}7
  rewrite addptE => -[_ zxy]. 7\label{line:no0}7
  exists [point of a0]; first exact: (@scaled_set_extract _ a).
  exists [point of b0]; first exact: scaled_set_extract.
  by eexists; rewrite zxy.
move/(congr1 (@S1 _)): zg; rewrite S1_convn. 7\label{line:hullmain}7
rewrite (bigID (fun i => g i \in X)) /=.
set b := \ssum_(i | _) _.
set c := \ssum_(i | _) _.
move=> zbc.
exists b; first exact: ssum_scaled_set.
exists c => //.
apply: (@ssum_scaled_set _ [pred i | g i \notin X]) => i /=.
move/asboolP; rewrite in_setE.
by case: (gT (g i) (imageP _ I)).
Qed.
\end{minted}

\section{Formalization of convex functions}
\label{sect:convex_functions}

In this section, we first (Sect.~\ref{sec:ordered_convex}) formalize a
generic definition of convex functions based on convex spaces; for
that purpose, we introduce in particular \newterm{ordered convex
  spaces}.
To demonstrate this formalization, we then apply it to the proof of
the concavity of the logarithm function and to an
information-theoretic function (Sect.~\ref{sec:convex_applications}).

\subsection{Ordered convex spaces and convex functions}
\label{sec:ordered_convex}

An ordered convex space extends a convex space with a partial order
structure:
\begin{definition}[
  \coqin{Module OrderedConvexSpace} in \cite{convexchoice}]
\label{def:ordered_convex_space}
An ordered convex space is a structure whose signature extends the one
of convex spaces as follows:
\begin{itemize}
\item Convex space $X$.
\item Ordering relation $(\_ \leq \_) \subset X \times X$.
\item Reflexivity law: $x \leq x$.
\item Transitivity law: $x \leq y \wedge y \leq z \imp x \leq z$.
\item Antisymmetry law: $x \leq y \wedge y \leq x \imp x = y$.
\end{itemize}
\end{definition}

The above definition does not force any interaction between convexity and
ordering.
It would also be a natural design to include an axiom stating that convex
combinations preserve ordering~\cite[Sect.~2]{keimel2016lmcs}.
We however do not need such interactions for defining convex functions, which is
our purpose here.

Convexity of a function is defined if its codomain is an ordered convex space.
In the following, let $T$ be a convex space and $U$ be an ordered convex space.

\begin{definition}
[\coqin{convex_function_at} in \cite{convexchoice}]
A~function $f : T \to U$ is convex at $p\in[0,1]$ and $x,y\in T$ if
$f(\pchoice x p y) \leq \pchoice {f(x)} p {f(y)}$.
\end{definition}

\begin{definition}
[\coqin{convex_function} in \cite{convexchoice}]
A~function $f : T \to U$ is convex if it is convex at all $p\in[0,1]$
and $x,y\in T$.
\end{definition}

The above predicates expect total functions. For partial functions,
we resort to convex sets (Def.~\ref{def:convex_subset}).

\begin{definition}
[\coqin{convex_function_in} in \cite{convexchoice}]
Let $D$ be a convex set in $T$.
A~function $f : T \to U$ is
convex in $D$ if it is convex at any $p\in[0,1]$ and $x,y\in D$.
\end{definition}

Concave functions are defined similarly since $f$ is concave for
the order $\leq$ if it is convex for $\geq$.
When the codomain of $f$ is $\mathbb{R}$, the prototypical example of
an ordered convex space, it is also easy to prove that $f$ is concave if $-f$ is
convex.

\subsection{Examples of convex functions}
\label{sec:convex_applications}

As a first example, we prove that the real logarithm function is concave.
The concavity of logarithm is frequently used in information theory, for
example, properties of data compression depend on it~\cite{affeldt2018isita}.

The definition of logarithm we use in \coq{} is the one of the
standard library; it has the entire $\mathbb R$ as its domain by
setting $\log(x)=0$ for $x\leq 0$.  The statement of concavity is then
restricted to the subset $\mathbb R_{>0}$.
\footnote{This way of restricting the domain of functions 
  in their properties rather than in the definitions is a design choice often found in
  \coq{}.  It makes it possible for functions such as the logarithm to be
  composable without being careful about their domains and ranges, and leads to a
  clean separation between definitions and properties of functions in the
  formalization.  }
\begin{lemma}[\coqin{log_concave} in \citefile{probability/ln\_facts.v}{infotheo}]
  The extended logarithm function
  \[
    x \mapsto
    \begin{cases}
      \log(x) & \text{if $x\in\mathbb R_{>0}$}\\
      0 & \text{otherwise}
    \end{cases}
  \]
  is concave in $\mathbb R_{>0}$.
\end{lemma}

The statement in \coq{} of these lemmas is as follows:
\begin{minted}{ssr}
Lemma log_concave : concave_function_in Rpos_interval log.
\end{minted}
The predicate \coqin{concave_function_in} has been explained
in Sect.~\ref{sec:ordered_convex}.
The object \coqin{Rpos_interval} is the set of positive numbers
described as the predicate \coqin{fun x => 0 < x} equipped with
the proof that this set is indeed convex.
The heart of the proof is the fact that a function whose second
derivative is non-negative is convex (\coqin{Section
  twice_derivable_convex} in \cite{convexchoice}).
Our proof proceeds by using the formalization of real analysis from
the \coq{} standard library; our formalization of convex spaces can
thus be seen as an added abstraction layer of convexity to this library.

\newcommand\dominates{<\!\!<}

\smallskip

Our second example of convex function is the \newterm{divergence}
(a.k.a.\ relative entropy or Kullback-Leibler divergence) of two
probability distributions: an important information-theoretic
function.
Let \coqin{P} and \coqin{Q} be two finite distributions (over some
finite type \coqin{A}). Their divergence \coqin{div} is defined as
follows:
\begin{minted}{ssr}
Variables (A : finType) (P Q : fdist A).
Definition div := \sum_(a in A) P a * log (P a / Q a).
\end{minted}
Actually, \coqin{div P Q} is defined only when \coqin{Q}
\newterm{dominates} \coqin{P}, i.e., when \coqin{Q a = 0} implies
\coqin{P a = 0} for all \coqin{a}. We call such a pair of probability
distributions a \newterm{dominated pair}.
Hereafter, we denote \coqin{div P Q} by \coqin{D(P || Q)} and the
dominance of \coqin{P} by \coqin{Q} by \coqin{P `<< Q}.

We now show that the divergence function is convex over the set of
dominated pairs. To formalize this statement using our definitions, we
first need to show that dominated pairs form a convex space. To
achieve this, it suffices to define the convex combination of the
dominated pairs \coqin{a `<< b} and \coqin{c `<< d} as \coqin{a <| p
  |> c `<< b <| p |> d} (where we use the convex combination of
probability distributions). This operator is easily shown to enjoy the
properties of convex spaces (Sect.~\ref{sect:convex_space}). Once this
is done, one just needs to uncurry the divergence function to use the
\coqin{convex_function} predicate:
\begin{minted}{ssr}
Lemma convex_div : convex_function (uncurry_dom_pair (@div A)).
\end{minted}
The proof follows the standard one~\cite[Thm.~2.7.2]{cover2006} and
relies on the log-sum inequality formalized in previous
work~\cite{affeldt2014jar}.

In previous work~\cite{affeldt2020cs}, we applied above results to the
proofs of convexity of other information-theoretic functions such as
the entropy and the mutual information.

\section{Related work}
\label{sec:related_work}

Conical spaces have been known in the literature to work as a nice-behaving
replacement of convex spaces when constructing models of nondeterministic
computations.
Varacca and Winskel~\cite{varacca2006mscs} used convexity when building a
categorical monad combining probability and nondeterminism, but they chose to
avoid the problem of equational laws in convex spaces by instead working with
conical spaces.
There is a similar preference in the study of domain-theoretic
semantics of nondeterminism, to a conical structure
(d-cones~\cite{kirch1993master}) over the corresponding convex
structure (abstract probabilistic domain~\cite{jones1989lics}).  The
problem is the same in this case: the difficulty in working with the
equational laws of convex spaces~\cite{keimel2009mscs, tix2009entcs}.

Flood~\cite{flood1981jams} proposed to use conical spaces to
investigate the properties of convex spaces.  He showed that for any convex
space, there is an enveloping conical space and the convex space is embedded in
it. (A version of the embedding for convex sets into cones in vector spaces was
already present in Semadini's book~\cite{semadini1971}.)
Keimel and Plotkin~\cite{keimel2016lmcs} extended the idea for their version of
ordered convex spaces and applied it in the proof of their key
lemma~\cite[Lemma~2.8]{keimel2016lmcs}, which is an ordered version of the one
proved by Neumann~\cite[Lemma 2]{neumann1970}.

Another aspect of convex spaces is the relationship to probabilistic
distributions.  From any set, one can freely generate a convex space
by formally taking all finite convex combinations of elements of this
set.
The resulting convex space can be seen as a set of distributions over
the original set, since the formal convex combinations are equivalent
to distributions over the given points.
By this construction, convex spaces serve as a foundation for the
algebraic and category-theoretic treatments of probability.  This
allows for another application of our work to the semantics of
probabilistic and nondeterministic programming~\cite{jacobs2010tcs,
  heerdt2018ictac}. We have also been investigating this
topic~\cite{monae,affeldt2019mpc}.  Our most recent
result~\cite{anonymous2020} is based on the properties of convex sets
and convex hulls, and deals with derived notions such as convex
powersets. Its purpose is the formal study of program semantics from a
category-theoretic point of view, rather than the formal study of the
mathematical structure of convex spaces itself, which is rather the
purpose of this paper.

\section{Conclusion}

In this paper, we formalized convex and conical spaces and developed
their theories.  In particular, we formally studied the various
presentations of the convex combination operator, be it binary or
multiary (Sect.~\ref{sect:convn}).
We provide formal proofs of the equivalence between several
axiomatizations of both operators, where ``proofs'' in the literature
were often only mere references to Stone's foundational
paper~\cite{stone1949annali}, while it only contains a reduction
of the multiary case to the binary one.
Based on convex and conical spaces, we
also developed a theory of convex functions and of convex hulls. We
illustrated these developments with detailed examples from real
analysis and information theory.

\paragraph{Acknowledgments}
We acknowledge the support of the JSPS KAKENHI Grant Number
18H03204. We also thank Shinya Katsumata for his comments.


\begin{thebibliography}{10}
\providecommand{\url}[1]{\texttt{#1}}
\providecommand{\urlprefix}{URL }
\providecommand{\doi}[1]{https://doi.org/#1}

\bibitem{cohen2018jfr}
Affeldt, R., Cohen, C., Rouhling, D.: Formalization techniques for asymptotic
  reasoning in classical analysis. J. Formaliz. Reason.  \textbf{11}(1),
  43--76 (2018)

\bibitem{anonymous2020}
Affeldt, R., Garrigue, J., Nowak, D., Saikawa, T.: A trustful monad for
  axiomatic reasoning with probability and nondeterminism (March 2020),
  \url{https://arxiv.org/abs/2003.09993}

\bibitem{monae}
Affeldt, R., Garrigue, J., Nowak, D., Saikawa, T., Sauvage, C., Tanaka, K.:
  Monadic equational reasoning in {Coq}.
  \url{https://github.com/affeldt-aist/monae/} (2019), {Coq} scripts.

\bibitem{affeldt2018isita}
Affeldt, R., Garrigue, J., Saikawa, T.: Examples of formal proofs about data
  compression. In: International Symposium on Information Theory and Its
  Applications (ISITA 2018), Singapore, October 28--31, 2018. pp. 665--669.
  IEICE. {IEEE} Xplore (Oct 2018)

\bibitem{affeldt2020cs}
Affeldt, R., Garrigue, J., Saikawa, T.: Reasoning with conditional
  probabilities and joint distributions in {Coq}. Computer Software  (2020),
  \url{https://staff.aist.go.jp/reynald.affeldt/documents/cproba_preprint.pdf},
  to appear. Japan Society for Software Science and Technology

\bibitem{affeldt2014jar}
Affeldt, R., Hagiwara, M., S\'enizergues, J.: Formalization of {Shannon}'s
  theorems. Journal of Automated Reasoning  \textbf{53}(1),  63--103 (2014)

\bibitem{affeldt2019mpc}
Affeldt, R., Nowak, D., Saikawa, T.: A hierarchy of monadic effects for program
  verification using equational reasoning. In: Hutton, G. (ed.) 13th
  International Conference on Mathematics of Program Construction (MPC 2019),
  7--9 October 2019, Porto, Portugal. pp. 226--254. Springer (2019)

\bibitem{beaulieu2008phd}
Beaulieu, G.: Probabilistic completion of nondeterministic models. Ph.D.
  thesis, University of Ottawa (2008)

\bibitem{bertot2008tphols}
Bertot, Y., Gonthier, G., Ould~Biha, S., Pasca, I.: Canonical big operators.
  In: Mohamed, O.A., Mu{\~{n}}oz, C., Tahar, S. (eds.) 21st International
  Conference on Theorem Proving in Higher Order Logics (TPHOLs 2008), Montreal,
  Canada, August 18-21, 2008. pp. 86--101. Springer (2008)

\bibitem{bonchi2017concur}
Bonchi, F., Silva, A., Sokolova, A.: {The Power of Convex Algebras}. In: Meyer,
  R., Nestmann, U. (eds.) 28th International Conference on Concurrency Theory
  (CONCUR 2017). Leibniz International Proceedings in Informatics (LIPIcs),
  vol.~85, pp. 23:1--23:18. Schloss Dagstuhl--Leibniz-Zentrum fuer Informatik
  (2017). \doi{10.4230/LIPIcs.CONCUR.2017.23}

\bibitem{cheung2017phd}
Cheung, K.H.: Distributive Interaction of Algebraic Effects. Ph.D. thesis,
  University of Oxford (2017)

\bibitem{cover2006}
Cover, T.M., Thomas, J.A.: Elements of information theory. Wiley (2006), 2nd
  ed.

\bibitem{flood1981jams}
Flood, J.: Semiconvex geometry. Journal of the Australian Mathematical Society
  \textbf{30}(4),  496–510 (1981). \doi{10.1017/S1446788700017973}

\bibitem{fritz2015arxiv}
Fritz, T.: Convex spaces {I}: Definition and examples (2015), available at
  \url{https://arxiv.org/abs/0903.5522}. First version: 2009

\bibitem{garillot2009tphols}
Garillot, F., Gonthier, G., Mahboubi, A., Rideau, L.: Packaging mathematical
  structures. In: 22nd Int.\ Conf.\ on Theorem Proving in Higher Order Logics
  (TPHOLs 2009), Munich, Germany, August 17--20, 2009. Lecture Notes in
  Computer Science, vol.~5674, pp. 327--342. Springer (2009)

\bibitem{heerdt2018ictac}
van Heerdt, G., Hsu, J., Ouaknine, J., Silva, A.: Convex language semantics for
  nondeterministic probabilistic automata. In: Fischer, B., Uustalu, T. (eds.)
  15th International Colloquim on Theoretical Aspects of Computing (ICTAC
  2018), 12--19 October 2018, Stellenbosch, South Africa. pp. 472--492.
  Springer (2018)

\bibitem{infotheo}
Infotheo: A {Coq} formalization of information theory and linear
  error-correcting codes. \url{https://github.com/affeldt-aist/infotheo/}
  (2020), {Coq} scripts.

\bibitem{convexchoice}
Infotheo: {\tt probability/convex\_choice.v}. In:  \cite{infotheo} (2020),
  {Coq} scripts.

\bibitem{jacobs2010tcs}
Jacobs, B.: Convexity, duality and effects. In: {IFIP} {TCS}. {IFIP} Advances
  in Information and Communication Technology, vol.~323, pp. 1--19. Springer
  (2010)

\bibitem{jones1989lics}
{Jones}, C., {Plotkin}, G.D.: A probabilistic powerdomain of evaluations. In:
  [1989] Proceedings. Fourth Annual Symposium on Logic in Computer Science. pp.
  186--195 (June 1989). \doi{10.1109/LICS.1989.39173}

\bibitem{keimel2016lmcs}
Keimel, K., Plotkin, G.: Mixed powerdomains for probability and nondeterminism.
  Logical Methods in Computer Science  \textbf{13} (12 2016).
  \doi{10.23638/LMCS-13(1:2)2017}

\bibitem{keimel2009mscs}
Keimel, K., Plotkin, G.D.: Predicate transformers for extended probability and
  non-determinism. Mathematical Structures in Computer Science  \textbf{19}(3),
   501–539 (2009). \doi{10.1017/S0960129509007555}

\bibitem{kirch1993master}
Kirch, O.: Bereiche und Bewertungen. Master's thesis, Technischen Hochschule
  Darmstadt (1993)

\bibitem{mahboubi2013itp}
Mahboubi, A., Tassi, E.: Canonical structures for the working {Coq} user. In:
  4th International Conference on Interactive Theorem Proving ({ITP} 2013),
  Rennes, France, July 22--26, 2013. Lecture Notes in Computer Science,
  vol.~7998, pp. 19--34. Springer (2013)

\bibitem{neumann1970}
Neumann, W.D.: On the quasivariety of convex subsets of affine spaces. Archiv
  der Mathematik  \textbf{21},  11--16 (1970)

\bibitem{semadini1971}
Semadini, Z.: Banach Spaces of Continuous Functions. PWN (1971)

\bibitem{stone1949annali}
Stone, M.H.: Postulates for the barycentric calculus. Ann. Mat. Pura Appl.
  \textbf{29}(1),  25--30 (1949)

\bibitem{swirszcz1974bapmam}
\'Swirszcz, T.: Monadic functors and convexity. Bulletin de l'Acad\'emie
  polonaise des sciences. S\'erie des sciences math\'ematiques, astronomiques
  et physiques  \textbf{22}(1) (1974)

\bibitem{coq}
{The Coq Development Team}: The {Coq} Proof Assistant Reference Manual. Inria
  (2019), available at \url{https://coq.inria.fr}. Version 8.11.0

\bibitem{tix2009entcs}
Tix, R., Keimel, K., Plotkin, G.: Semantic domains for combining probability
  and non-determinism. Electronic Notes in Theoretical Computer Science
  \textbf{222},  3 -- 99 (2009).
  \doi{https://doi.org/10.1016/j.entcs.2009.01.002}

\bibitem{varacca2006mscs}
Varacca, D., Winskel, G.: Distributing probability over non-determinism.
  Mathematical Structures in Computer Science  \textbf{16}(1),  87--113 (2006)

\end{thebibliography}

\end{document}